\documentclass[showpacs,preprintnumbers,twocolumn,amsmath,amssymb,pre]{revtex4}
\usepackage[english]{babel}
\usepackage[utf8]{inputenc}
\usepackage[T1]{fontenc}

\usepackage{graphicx}
\usepackage{dcolumn}
\usepackage{bm}
\usepackage{enumerate}

\begin{document}

\title{How to determine a boundary condition at a thin membrane for diffusion from experimental data}

\author{Tadeusz Koszto{\l}owicz}
 \email{tadeusz.kosztolowicz@ujk.edu.pl}
 \affiliation{Institute of Physics, Jan Kochanowski University,\\
         ul. \'Swi\c{e}tokrzyska 15, 25-406 Kielce, Poland}
         
\author{Katarzyna D. Lewandowska}
 \email{kale@gumed.edu.pl}
 \affiliation{Department of Radiological Informatics and Statistics, Medical University of Gda\'nsk,\\ 
 ul. Tuwima 15, 80-210 Gda\'nsk, Poland}

\author{S{\l}awomir W\c{a}sik}
 \email{slawomir.wasik@ujk.edu.pl}
 \affiliation{Institute of Physics, Jan Kochanowski University,\\
         ul. \'Swi\c{e}tokrzyska 15, 25-406 Kielce, Poland}

\date{\today}

\begin{abstract}
We present a new method of deriving a boundary condition at a thin membrane for diffusion from experimental data. Based on experimental results obtained for normal diffusion of ethanol in water, we show that the derived boundary condition at a membrane contains a term with the Riemann--Liouville fractional time derivative of the $1/2$ order. Such a form of the  boundary condition shows that a transfer of particles through a thin membrane is a `long memory process'. Presented method is an example that an important part of mathematical model of physical process may be derived directly from experimental data.

\end{abstract}


\maketitle

Normal diffusion in a system with a thin membrane is considered in many problems of substances transport occurring in life sciences and engineering. The bibliography of this subject is lengthy; we mention here the books \cite{hobbie,luckey,hsieh}. In order to solve the diffusion equation, which describes this process, two boundary conditions at a membrane are needed. Usually, one of them supposes a continuity of flux at a membrane. Within the commonly used method of modelling diffusion in a membrane system, the form of the second boundary condition is assumed. Unfortunately, this condition can been chosen in various forms which are nonequivalent. In contrast to such a method, we derive a boundary condition at a thin membrane directly from experimental data; any assumption about the detailed form of the boundary condition is not needed. 

We study both experimentally and theoretically normal diffusion of ethanol in water in a system which is divided into two parts by a thin membrane which is localized at $x=0$. We assume that the system under study is homogeneous in a plane perpendicular to the $x$-axis thus, the system is effectively one--dimensional and that diffusion is described by the equation $\partial C(x,t)/\partial t=D\partial^2 C(x,t)/\partial x^2$ with a constant diffusion coefficient $D$; $C(x,t)$ denotes particles concentration.
We suppose that diffusing particles are not accumulated inside a membrane and a membrane permeability does not depend on the concentration. In the following, functions describing the process in the region $x<0$ are marked by the subscript $1$ and in the region $x>0$ by the subscript $2$. 
We assume that the first boundary condition requires a continuous flux at the membrane
$J_1(0^-,t)=J_2(0^+,t)$, where $J_{1,2}(x,t)=-D\partial C_{1,2}(x,t)/\partial x$.
We find the second boundary condition on the basis of experimentally obtained concentration profiles. 
We conduct our consideration in terms of the Laplace transform $\mathcal{L}\{f(t)\}\equiv\hat{f}(p)=\int_{0}^{\infty}{\rm e}^{-pt}f(t)dt$.
Let us assume the second boundary condition in the form
\begin{equation}
  \label{eq1}
  \hat{C}_2(0^+,p)=\hat{\Phi}(p)\hat{C}_1(0^-,p)\;,
\end{equation}
where $\hat{\Phi}(p)$ is a function to be determined. Since we suppose that particles move independently and do not clog a membrane, $\Phi(t)$ is independent of particles concentration. 
The inverse Laplace transform of the boundary condition~(\ref{eq1}) is
\begin{equation}
  \label{eq2}
  C_2(0^+,t)=\int_{0}^{t}\Phi(t-t')C_1(0^-,t')dt'\;.
\end{equation}
Although the presented model assumes that a membrane is infinitely thin, we will show that the obtained results describe diffusion well also in a system with a thin membrane which has a finite thickness.

If $\hat{\Phi}(p)=\kappa$, where $\kappa$ is a positive constant, then $\Phi(t)=\kappa\delta(t)$, where $\delta$ denotes the Dirac--delta function, and we call the boundary condition (\ref{eq2}) `memoryless'. We note that many others boundary conditions given in terms of the Laplace transform can be expressed by Eq.~(\ref{eq1}). Some of the examples are as follows: for a boundary condition, which is sometimes called the radiation or the Robin boundary condition $J(0^-,t)=\lambda C_1(0^-,t)$, $\lambda>0$, we obtain $\hat{\Phi}(p)=1/(\lambda^2-1+\lambda\sqrt{Dp})$. For $J(0,t)=\gamma\left[C_1(0^-,t)-C_2(0^+,t)\right]$, $\gamma>0$, we have $\hat{\Phi}(p)=1/(1+(\sqrt{Dp}/\gamma))$. For a fully reflecting wall we get $\hat{\Phi}(p)\equiv 0$ whereas a fully absorbing wall corresponds to $\hat{\Phi}(p)\equiv \infty$. If the appearance of particles on the right surface of a membrane is delayed in time by $\tau$ with respect to their appearance on the left membrane surface then $\hat{\Phi}(p)={\rm e}^{-\tau p}$. Above examples show that Eq.~(\ref{eq1}) can be treated as the general form of a boundary condition at a membrane under assumptions taken in our considerations. 

Since $\Phi(t)$ is assumed to be independent of a concentration, we choose the initial concentration in a form which is convenient for experimental investigation, namely,
\begin{equation}
  \label{eq3}
  C(x,0)=\left\{
  \begin{array}{ll}
    C_0, & x<0,\\
    0, & x>0.
  \end{array}
  \right.
\end{equation}
Then, the Laplace transforms of solutions to the diffusion equation read 
\begin{equation}
  \label{eq4}
  \hat{C}_1(x,p)=\frac{C_0}{p}-\frac{C_0\hat{\Phi}(p)}{p\left[\hat{\Phi}(p)+1\right]}{\rm e}^{x\sqrt{\frac{p}{D}}}\;,
\end{equation}
\begin{equation}
  \label{eq5}
  \hat{C}_2(x,p)=\frac{C_0\hat{\Phi}(p)}{p\left[\hat{\Phi}(p)+1\right]}{\rm e}^{-x\sqrt{\frac{p}{D}}}\;.
\end{equation}

It is convenient to find a function $\hat{\Phi}(p)$ on the base of a time evolution of an amount of substance which passes a membrane $W(t)=A\int_{0}^{\infty}C_2(x,t)dx$, where $A$ is the cross sectional area of a vessel in which diffusion occurs. Taking into account Eq.~(\ref{eq5}) we get
\begin{equation}
  \label{eq6}
  \frac{\hat{W}(p)}{A}=\frac{C_0\sqrt{D}\hat{\Phi}(p)}{p^{3/2}\left[\hat{\Phi}(p)+1\right]}\;.
\end{equation}

The experimental measurement of concentration profiles has been conducted in a membrane system which is a vessel consists of two glass cuboid-like shaped cuvettes ($7$ mm wide, $10$ mm high, $65$ mm long) separated by a horizontally located thin membrane; the area of the membrane surface is $A=70$ mm$^2$. Initially, the upper cuvette ($x<0$) was filled with an aqueous solution of ethanol with $C_0=0.250\times10^{-6}$ mol/mm$^3$ whereas, the lower cuvette ($x>0$) was filled with pure water. Since a concentration gradient is in the vertical direction alone, diffusion is effectively one-dimensional. A substance concentration is measured by means of the laser interferometric method; the method and the setup used in the experiment are described in \cite{kdm,wadkg}. Because of technical reasons the measurement of a concentration has only been conducted in the lower vessel up to $6600$ seconds. The artificial nephrophan hemodialyzer membrane, $(16\pm 2)\times 10^{-3}$ mm in thickness, was made of cellulose acetate. The measurements were performed at $293$ K. The diffusion coefficient of ethanol in water solvent, $D=(0.95\pm 0.12)\times 10^{-3}$ mm$^2$/s, was experimentally determined by means of the time evolution of near membrane layers method which is described in \cite{kdm}. Although the membrane is horizontal, the gravitational effect is negligibly small and concentration of ethanol is described with a good accuracy by solution to the diffusion equation without a migration term.

Ten independent measurement series of concentration profiles have been performed. Concentration profiles were recorded every $120$ s in time interval $(0,2400)$ s and every $600$ s in $(2400, 6600)$ s. Having ten values of $C_2(x,t)$ we calculated the mean
value of concentration and the standard deviation for given $x$ and $t$. The function $W(t)/A$ was calculated by means of the numerical integration of concentrations. The errors shown in Figs. \ref{fig2} and \ref{fig4} were obtained by further multiplying the standard deviations by the Student-Fisher coefficient taken at a confidence level $95\%$ to include the effect of low statistics.

\begin{figure}[h!]
  \centering
  \includegraphics[scale=0.25, angle=-90.0]{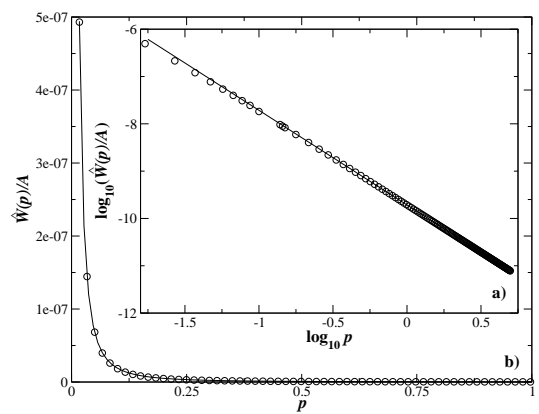}
  \caption{Plot {\rm a}: The function $\hat{W}(p)$ obtained numerically from the experimental data (symbols) in the logarithmic scale, a solid line represents the function ${\rm log_{10}}(\hat{W}(p)/A)=-2{\rm log_{10}} p-9.76$. Plot {\rm b}: $\hat{W}(p)$ obtained numerically from the experimental data (symbols), solid line represents Eq. (\ref{eq7}) with $a=2.89\times 10^8$ mm$^2\sqrt{\rm s}$/mol and $b=10^{9.76}$ mm$^2\;$s/mol. \label{fig1}}
\end{figure}

\begin{figure}[h!]
  \centering
  \includegraphics[scale=0.25,angle=-90]{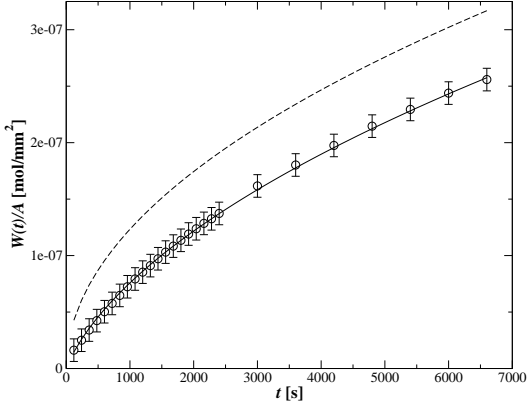}
  \caption{The plot of $W(t)/A$ versus time, symbols represent the experimental data, the solid line represents the function (\ref{eq11}), and the dashed line the function (\ref{eq14}), the plots of theoretical functions are obtained for $C_0=0.250\times 10^{-6}$ mol/mm$^3$, $D=9.5\times 10^{-4}$ mm$^2$/s, $\alpha=1.23$, and $\beta=44.12\;\sqrt{s}$. \label{fig2}}
\end{figure}

\begin{figure}[h!]
  \centering
  \includegraphics[scale=0.25,angle=-90]{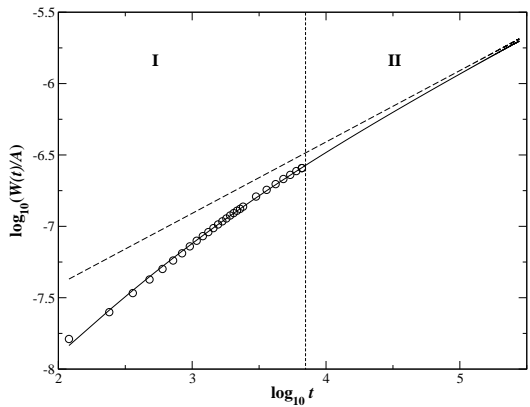}
  \caption{The plot of $W(t)/A$ versus time in the logarithmic scale; the description is similar to Fig. \ref{fig2}. The vertical dotted line shows the limit of time interval, $t=6600$ s, in which measurements were made. The region I corresponds to the plots presented in Fig. \ref{fig2}, in the region II there are shown theoretical functions only; it seems that treatment of the solid line as an extrapolation of the experimental results in the region II is well motivated.  \label{fig3}}
\end{figure}

\begin{figure}[h!]
\centering
\includegraphics[scale=0.25,angle=-90.0]{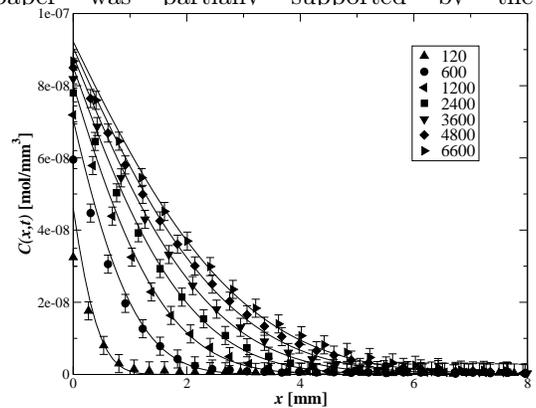}
\caption{Concentration profiles for different times, given in seconds, presented in the legend. The parameters are the same as in Fig. \ref{fig2}. Symbols represent the experimental data, solid lines represents the function (\ref{eq10}). \label{fig4}}
\end{figure}

Below we will compare the theoretical function~(\ref{eq5}) with numerically calculated Laplace transform of $W(t)/A$ obtained from experimental concentration profiles. Numerical calculations have been performed by means of the Gauss--Laguerre quadrature and the spline interpolation method. The result is presented in Fig.~\ref{fig1}.
The function $\hat{W}(p)/A$ has the following assymptotic properties. In Fig. \ref{fig1} there is observed $\hat{W}(p)/A\approx 1/bp^2$ for $p\rightarrow \infty$. In Fig. \ref{fig3} we observe $W(t)/A\sim t^{1/2}$ in the long time limit. Since the limit of long time corresponds to the limit of small $p$ and $\mathcal{L}\{t^{1/2}\}=\Gamma(3/2)/p^{3/2}$, we deduce that $\hat{W}(p)/A\approx 1/ap^{3/2}$ for $p\rightarrow 0$. Guided by the properties mentioned above we suppose that
\begin{equation}\label{eq7}
\frac{\hat{W}(p)}{A}=\frac{1}{ap^{3/2}+bp^2}\;,
\end{equation} 
where $a$ and $b$ are parameters to be determined.
From Fig.~\ref{fig1} we have $\hat{W}(p)/A=10^{-9.76}/p^2$ for large $p$, thus we get $b=10^{9.76}$ mm$^2\;$s/mol. The parameter $a$ ensures the best fit of the function (\ref{eq7}) to the empirical data in the whole domain of $p$ and we obtain $a=2.89\times 10^8$ mm$^2\sqrt{\rm s}$/mol.

Equations (\ref{eq6}) and (\ref{eq7}) provide
\begin{equation}\label{eq8}
\hat{\Phi}(p)=\frac{1}{\alpha+\beta\sqrt{p}},
\end{equation}
where $\alpha=C_0\sqrt{D}a-1$ and $\beta=C_0\sqrt{D}b$; $\alpha$ and $\beta$ are independent of $C$ and $D$ and control a membrane permeability. Taking the values of parameters presented earlier we get $\alpha=1.23$ and $\beta=44.12\;\sqrt{\rm s}$. 
For $\beta\neq 0$ we get
\begin{equation}\label{eq9}
\Phi(t)=\frac{1}{\beta}\left[\frac{1}{\sqrt{\pi t}}-\frac{\alpha}{\beta}\;{\rm e}^{\frac{\alpha^2 t}{\beta^2}}{\rm erfc}\left(\frac{\alpha\sqrt{t}}{\beta}\right)\right]\;, 
\end{equation}
where ${\rm erfc}(u)\equiv(2/\sqrt{\pi})\int_u^\infty{\rm e}^{-\xi^2}d\xi$ is the complementary error function, and
\begin{eqnarray}\label{eq10}
\lefteqn{C_2(x,t)=\frac{C_0}{1+\alpha}\;{\rm erfc}\left(\frac{x}{2\sqrt{Dt}}\right)}\nonumber\\
&&-\frac{C_0}{1+\alpha}\;{\rm e}^{\frac{(1+\alpha)x}{\beta\sqrt{D}}+\frac{(1+\alpha)^2 t}{\beta^2}}\nonumber\\
&&\mbox{} \times{\rm erfc}\left(\frac{x}{2\sqrt{Dt}}+\frac{(1+\alpha)\sqrt{t}}{\beta}\right)\;,
\end{eqnarray}
\begin{eqnarray}\label{eq11}
\lefteqn{\frac{W(t)}{A}=\frac{2C_0\sqrt{Dt}}{(1+\alpha)\sqrt{\pi}}-\frac{C_0\sqrt{D}\beta}{(1+\alpha)^2}}\nonumber\\
&&+\;\frac{C_0\sqrt{D}\beta}{(1+\alpha)^2}\;{\rm e}^{\frac{(1+\alpha)^2 t}{\beta^2}}\;{\rm erfc}\left(\frac{(1+\alpha)\sqrt{t}}{\beta}\right)\;.
\end{eqnarray}

The boundary condition at a membrane is determined by Eqs. (\ref{eq2}) and (\ref{eq9}). This condition can be written in the other form. Namely, from Eqs. (\ref{eq1}) and (\ref{eq8}) we get $\alpha \hat{C}_2(0^+,p)+\beta\sqrt{p}\hat{C}_2(0^+,p)=\hat{C}_1(0^-,p)$. This equation and the relation $\mathcal{L}^{-1}\left\{\sqrt{p}\hat{f}(p)\right\}=d^{1/2}f(t)/dt^{1/2}$, where $d^{1/2}f(t)/dt^{1/2}=(1/\sqrt{\pi})\left(d/dt\right)\int_0^tdt'f(t')/(t-t')^{1/2}$ denotes the Riemann--Liouville fractional derivative of the order $1/2$, provide 
\begin{equation}
  \label{eq12}
\alpha C_2(0^+,t)+\beta\frac{\partial^{1/2}}{\partial t^{1/2}}\;C_2(0^+,t)=C_1(0^-,t)\;.
\end{equation}

For $\beta=0$ we get the memoryless process for which $\Phi(t)=\delta(t)/\alpha$. In this case we obtain
\begin{equation}\label{eq13}
C_{2}(x,t)=\frac{C_0}{1+\alpha}\;{\rm erfc}\left(\frac{x}{2\sqrt{Dt}}\right),
\end{equation}
and
\begin{equation}\label{eq14}
\frac{W(t)}{A}=\frac{2C_0\sqrt{D}}{\sqrt{\pi}(1+\alpha)}\sqrt{t}\;.
\end{equation}
We note that Eqs. (\ref{eq10}) and (\ref{eq11}) take the form of Eqs. (\ref{eq13}) and (\ref{eq14}), respectively, in the limit of long time. For this reason, $\alpha$, which is independent of $\beta$, is assumed to be the same in Eqs. (\ref{eq8})--(\ref{eq14}). In Figs. \ref{fig2}--\ref{fig4} the functions (\ref{eq10}) and (\ref{eq11}) are in a good agreement with the experimental data for the parameters presented in the text, in contrast to the function (\ref{eq14}) which corresponds to the `memoryless' boundary condition. 

The presence of a fractional derivative in the boundary condition (\ref{eq12}) shows that passing of particles through a thin membrane is a long memory process. 
This is astonishing because normal diffusion is usually considered as the Markovian proces or possibly as a process with a short memory if the Cattaneo hyperbolic diffusion equation is considered. In Eq.~(\ref{eq12}) the term generating long memory effect vanishes over time. The question arises what is the approximate time after that this term can be omitted. 
`Memory length' of the process is also manifested in $\Phi(t)$. For short time the function (\ref{eq9}) takes the form $\Phi(t)\approx 1/\beta\sqrt{\pi t}$. In this case the kernel of integral operator in Eq. (\ref{eq2}) is the same as in the Riemann--Liouville derivative of the order $1/2$ and the function $\Phi(t)$ creates a long--memory process. For long time, using the relation $\sqrt{\pi}{\rm e}^{u^2}{\rm erfc}\;u\approx 1/u-1/2u^3$, $u\rightarrow\infty$, we get $\Phi(t)\approx (\beta/2\sqrt{\pi}\alpha^2)t^{-3/2}$; this function generates a process with a relative short memory.

The fundamental solution to the diffusion equation $P(x,t;x_0)$ for which $P(x,0;x_0)=\delta(x-x_0)$, can be interpreted as a probability density of finding a particle at point $x$ at time $t$; $x_0$ is a particle's location for $t=0$. 
Since $P_2(0^+,t;x_0)=\int_{0}^{t}\Phi(t-t')P_1(0^-,t';x_0)dt'$, the interpretation of Eq. (\ref{eq2}) is that a probability of finding a particle at the membrane surface $x=0^+$ depends on the history of its appearance on the opposite membrane surface. Let us denote by $M_1(0^-,t;x_0)$ and $M_2(0^+,t;x_0)$ a mean number of visits of a particle at the membrane surface $x=0^-$ and $x=0^+$, respectively. Since $\hat{M}(0^\pm,s;x_0)=\hat{P}(0^\pm,s;x_0)/(1-\hat{\omega}(s))$ \cite{montroll65}, where $\omega(t)$ is a probability density of time which is needed to take particle's next step, the following relation is valid
\begin{equation}\label{eq15}
M_2(0^+,t;x_0)=\int_{0}^{t}\Phi(t-t')M_1(0^-,t';x_0)dt\;.
\end{equation}
Thus, the mean number of visits at the membrane surface $x=0^+$ in time interval $(0,t)$ depends on the long history of mean number of visits at the opposite membrane surface. A simple stochastic model of diffusion in a membrane system, which provides a boundary condition similar to Eq. (\ref{eq12}) in which the order of fractional derivative depends on a kind of diffusion, is presented in \cite{tk}.

In summary, the most important results presented in the paper and conclusions are as follows: (i) we have presented the new method of deriving a boundary condition at a thin membrane from experimental data. Within this method the Laplace transform of a boundary condition is assumed to be in the form of Eq. (\ref{eq1}). Next, one finds the Laplace transform of some theoretical function contained $\Phi$ which is relatively easy to experimental measurement. Then, this function is determined by means of numerical calculation of the Laplace transform of experimental data. Finally, comparing both Laplace transforms mentioned above one finds the function $\Phi$. (ii) The obtained boundary condition (\ref{eq12}) contains a term with a fractional order derivative. This term vanishes over time, but for ethanol diffusion in water in a system with nephrophan membrane it can be neglected after long time of the order of $10^5$ s, see Fig. \ref{fig3}. (iii) In our study we have assumed that an accumulation of particles inside a thin membrane does not occur. However, the presented method can be extended for the case of a thin or thick membrane in which molecules can be accumulated. Then, the boundary condition Eq. (\ref{eq1}) should be complemented by a second boundary condition such as $\hat{J}_2(0^+,s)=\hat{\Xi}(s) \hat{J}_1(0^-,s)$. To determine functions $\Phi$ and $\Xi$, concentrations profiles should be measured in both parts of the system. 

This paper was partially supported by the Polish National Science Centre under grant No. 2014/13/D/ST2/03608.

\end{document}